\documentclass[prl,twocolumn,nofootinbib,longbibliography]{revtex4-1}
\usepackage{filecontents}
\usepackage[resetlabels,labeled]{multibib}
\newcites{supp}{Supplementary References}
\usepackage{graphicx}
\usepackage{amssymb}
\usepackage{multibib}
\usepackage{color}
\usepackage{epsfig}
\usepackage{amsmath}
\usepackage[caption=false]{subfig}
\usepackage[colorlinks= true]{hyperref}
\hypersetup{colorlinks=true,citecolor=blue,}
\usepackage{lmodern}  
\usepackage[T1]{fontenc}

\begin{document}

\title{Kinetic Fragility Directly Correlates with the Many-body Static Amorphous Order in Glass-Forming Liquids}
\author{Indrajit Tah$^{1,2}$}
\email{itah@sas.upenn.edu}
\author{Smarajit Karmakar$^{2}$}
\email{smarajit@tifrh.res.in}
\affiliation{$^1$ Department of Physics and Astronomy,
University of Pennsylvania 209 South 33rd Street
Philadelphia, PA, USA }
\affiliation{$^2$ Tata Institute of Fundamental Research, 36/P, 
Gopanpally Village, Serilingampally Mandal,Ranga Reddy District, 
Hyderabad, 500107, India}

\begin{abstract}
The term ``fragility'' describes the rate at which viscosity grows when a supercooled liquid approaches its putative glass transition temperature. The field of glassy materials is actively searching for a structural origin that governs this dynamical slowing down in the supercooled liquid, which occurs without any discernible change in structure. Our work shows clear evidence that growing many-body static amorphous order is intimately correlated with the kinetic fragility of glass-forming liquids. It confirms that the system's dynamical response to temperature is concealed in its micro-structures. This finding may pave the way for a deeper understanding of the different temperature dependence of the relaxation time or viscosity in a wide variety of glass-forming liquids.
\end{abstract}

\maketitle
\noindent{\bf Introduction:}
The dramatic rise in viscosity or relaxation time upon supercooling is a universal hallmark feature across
all glass-forming liquids. Despite extensive investigations \cite{Berthier1797,RevModPhys.83.587,doi:10.1146/annurev-conmatphys-031113-133848,PhysRevLett.121.085703, doi:10.1063/1.5033555,Tahacsomega}, one of the fundamentally unsolved challenges in condensed matter physics is understanding of microscopic origin of rapid rise in viscosity ($\eta$) with relatively small temperature changes while approaching the calorimetric glass transition temperature ($T_g$), defined as the temperature at which $\eta$ becomes $10^{14} Poise$. In this context, it's also worth noting that, while near-diverging growth of viscosity is universal in all glass-forming liquids, the rate at which viscosity grows at low temperatures is quite non-universal and varies significantly across liquids. 

The term ``fragility'' was first coined in Ref.~\cite{Angell1924} to characterize this rapid non-universal changes in viscosity near $T_g$. Although, the word ``fragility'' is used to describe the dynamical properties, several theoretical and experimental studies \cite{sastry_nature,Martineznature,Kaori_nature,Debenedetti,Mauronatcomm,doi:10.1063/1.4964362} demonstrate that the fragility is fundamentally connected to thermodynamic properties of the liquid, like the excess entropy and the specific heat which are directly linked with the microscopic structure of a liquid. This connection led to a search for the structural or thermodynamic correlations in these amorphous systems, which may qualitatively describe the source of fragility. However, the correlation between thermodynamics and dynamical properties in supercooled liquids remains somewhat controversial, leading to inherent uncertainty, which prevents researchers from reaching a consensus. Thus,  to establish a definitive correlation between thermodynamic properties and kinetic fragility further research will be welcomed. The outcome will certainly have significant implications in understanding some of these puzzles in the dynamics of supercooled liquids. In this article, we have addressed the connection between growing static amorphous order and fragility by performing extensive molecular dynamics simulations of a model glass-forming liquid, showing very large variation in fragility with increasing density.

The temperature dependence of viscosity and relaxation time for a wide variety of supercooled liquids
can be well fitted by the Vogel-Fulcher-Tamann (VFT) formula \cite{10004192038,https://doi.org/10.1111/j.1151-2916.1925.tb16731.x,https://doi.org/10.1002/zaac.19261560121}:
\begin{equation}
\tau_\alpha(T)=\tau_0 \exp\left [\frac{1}{K_{VFT} \ (T/T_{VFT}-1)}\right],
\end{equation}
where $\tau_\alpha$ is the structural relaxation time (defined later), $\tau_0$ is the viscosity at 
infinite high temperature, $T_{VFT}$ is the apparent divergence temperature for relaxation time and 
$K_{VFT}$ denotes the ``Kinetic fragility''. The fragility index provides a unifying framework for 
the classification of a broad range of systems, from molecular liquids \cite{Angell1924} to 
colloidal  \cite{Frag_Molecu} and biological systems \cite{Sussman_2018,D0SM01575J}. Additionally, 
material functionality and manufacturability are directly linked with their fragility index. Like the 
system that have low fragility index, generally implies to have a wide glass transition temperature 
range which enhances the flexibility of material moulding. Moreover, material bulk properties depend 
on its molecular mechanisms and an important question is how the molecular mechanisms of 
glass forming liquids differ from each other in such a way that their dynamic properties vary so 
widely. Fragility also plays important role in bio-preservation\cite{Fox1922,https://doi.org/10.1111/j.1365-2621.1991.tb07970.x,doi:10.1063/1.5085077}. Empirical evidence suggests 
that the larger the fragility of a liquid, the better it will be in preserving biomacromolecules when 
used as a medium. Although, there are deviations from this hypothesis but this remained
as a rule of thumb in bio-preservation industry. Thus, a better understanding of fragility may 
lead to better understanding of the physics of biopreservation. 

\vskip +0.1in
\noindent{\bf Models and Methods:}
To gain valuable insights into molecular mechanisms and explore the connection between 
length scales associated with liquid structure and fragility, we have performed extensive computer 
simulations of soft repulsive particles \cite{PhysRevLett.88.075507,doi:10.1146/annurev-conmatphys-070909-104045,Berthier_2009} by varying the density, $(\rho)$,  and temperature, 
$(T)$, which cover a  broad spectrum of fragility. This model (referred as HP model) shows a 
strong crossover from strong glass to fragile glass behaviour with varying density $(\rho)$ beyond 
the jamming density $(\rho_J)$ \cite{Berthier_2009}. This provides a conceptual and more feasible 
way to tune fragility  without changing the particle composition, interaction potential or curving the 
configurational space \cite{PhysRevLett.101.155701}. Simulations are done in three-dimensions in 
the density range $\rho \in [0.65, 0.82]$ with $N = 108000$ particles. 
More detailed information about the models and simulations is provided in SI \cite{SI}.
To describe the dynamics of the system, we have computed the average relaxation timescale, 
$\tau_{\alpha}$, of the system via  the two-point overlap correlation function $(Q(t))$ (defined in the 
SI \cite{SI}) at each equilibrium density, $\rho$, and temperature, $T$ state points. The  relaxation 
time $\tau_\alpha$ is defined as $\langle Q(\tau_\alpha)\rangle = 1/e$, where $\langle\cdots\rangle$ 
refers to ensemble average. The calorimetric glass transition temperature, $T_g$ for this model in
simulations is defined as $\tau_\alpha(T_g) = 5 \times 10^6$.

\vskip +0.05in
\noindent{\bf Results: }
\begin{figure}[htp]
\includegraphics[scale=0.22]{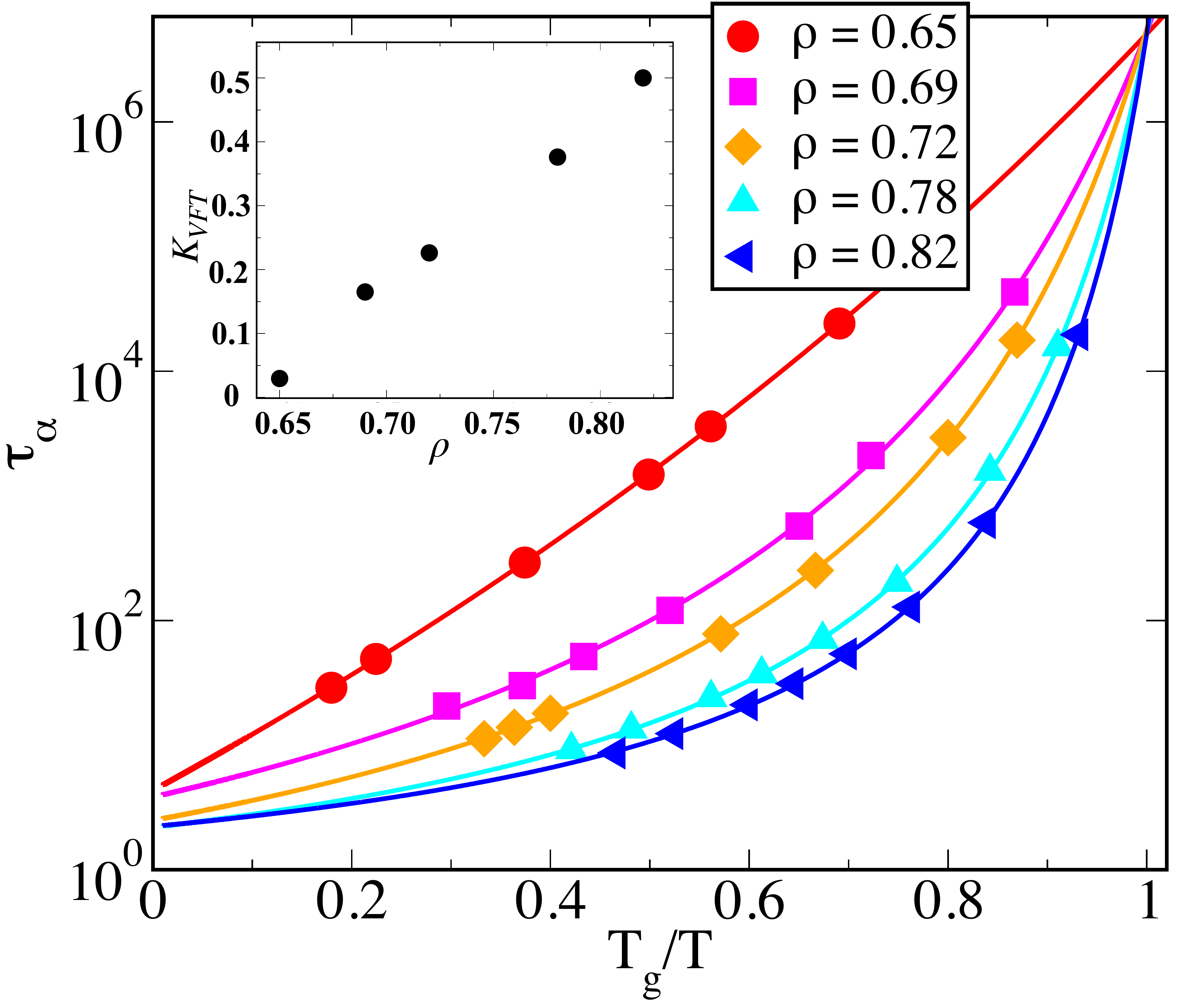}
\caption{Angell plot of $(\tau_\alpha)$ as a function of rescaled temperature, $T_g/T$ at different 
densities ($\rho$). At higher density  the relaxation time exhibits a sharp growth upon decreasing 
temperature, a feature of a fragile glass-former, while at lower density the relaxation time displays 
Arrhenius behaviour with temperature, suggesting a crossover to strong glass-former.  Lines are
the fit to VFT form. Inset shows the  fragility index against density.}
  \label{fig:1}
\end{figure}
Fig.~\ref{fig:1} shows the Angell plot of the relaxation time $(\tau_\alpha)$ as a function  of rescaled 
inverse temperature. To have a fair comparison, the data has been normalized by respective $T_g$ 
at different density. At low density the relaxation time exhibits 
Arrhenius dependence, indicating a behaviour of a strong glass forming liquid, whereas at high density 
the relaxation time display super-Arrhenius  temperature dependence, often referred as 
``Fragile glass-former''. In these models, fragility can be tuned over a broad range, by only increasing 
the density without changing any other microscopic characteristics of the systems. This gives us a big
opportunity to look into the precise molecular mechanisms that cause such large change in fragility. 
In the inset of Fig.~\ref{fig:1}, we plot the fragility index $(K_{VFT})$ against density. One can see 
that increasing bulk density by a factor of $1.26$ ($\rho = 0.65$ to $0.82$) changes the 
kinetic fragility by a factor of $17$ ($K_{VFT} = 0.0295$ to $0.5002$). Thus this model
is an ideal test-bed for deciphering the microscopic origin of fragility and its possible connection 
to growing amorphous order as envisaged in Random First Order Transition (RFOT) Theory \cite{PhysRevA.40.1045,doi:10.1146/annurev.physchem.58.032806.104653}.
In the subsequent paragraphs, we will discuss the strong correlation between changing fragility 
and the two important growing length scales in the systems; namely dynamic heterogeneity
length scale ($\xi_d$) and static amorphous order length scale ($\xi_s$).

\vskip +0.05in
\noindent{\bf Dynamic Length Scale:}
Since, fragility is measured from dynamical properties like relaxation time, it is natural to 
investigate the temperature dependence of $\xi_d$ for these model systems and try to understand
a possible correlation between them. Recent work \cite{doi:10.1063/1.4938082} pointed out 
that growth of dynamic heterogeneity (DH) in strong liquids is different than in fragile liquids. Thus, a 
systematic study of DH \cite{doi:10.1146/annurev.physchem.51.1.99} in this current set up is indeed 
warranted to have a detailed understanding of 
how heterogeneity gets affected by changing fragility in the system. 
DH  can be quantified via the fluctuation of two-point 
correlation function (Q(t)) \cite{PhysRevE.66.030101} and the length scale associated with DH, can 
be measured from the spatial correlation of four-point structure factor (fluctuation of two point correlation 
function in fourier space, $S_4(q,t=\tau_\alpha)$) \cite{PhysRevE.58.3515,doi:10.1063/1.1605094}.
In SI \cite{SI}, the determination of $\xi_d$ from $S_4(q,t=\tau_\alpha)$ has been elaborated. 
Recently, in Ref.~\cite{PhysRevLett.119.205502}, another method of extracting dynamic length 
scale has been proposed. In this method the dynamical properties of the systems is measured at 
a smaller sub volume of the systems by diving the whole systems in to smaller blocks of length, $L_B$.
This method can be easily implemented both in experiments and in 
numerical simulations and termed as ``block analysis'' method. The method has also been shown 
to significantly improve the statistical averaging of the data as well as include all possible fluctuations 
({\it e.g} density, temperature, composition, {\it etc.}) that are important in measuring four-point correlation 
functions (see SI \cite{SI} for further details). We have obtained the dynamic length scales using both
the methods for reliability.

\begin{figure}[htbp]
\includegraphics[scale=0.26]{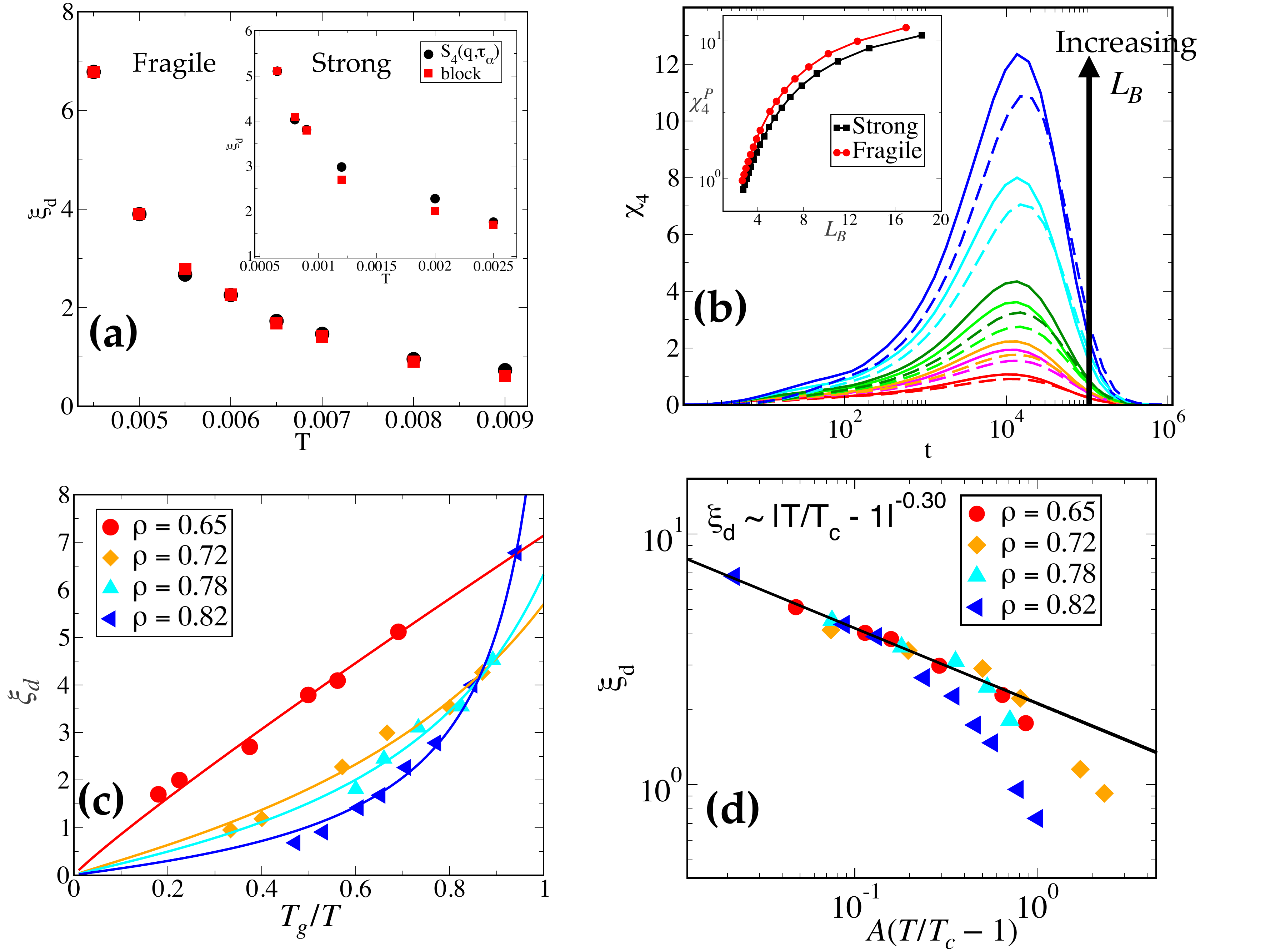}
\caption{{\textbf{(a).} Dynamics length scale as a function of temperature for fragile liquid. The legend shows the method of extraction of length scale. Similar plot for strong glass forming liquid is shown in the inset. \textbf{(b)}. Dynamical heterogeneity as a function of time for various block size ($L_B$). Fragile liquids ($\rho = 0.82$, solid lines), shows stronger growth of dynamic heterogeneity compare to  strong liquids ($\rho=0.65$, 
dotted lines). We display $\chi_4(t)$ for the strong and fragile liquids at roughly the same block scale. Fragile 
liquids display more heterogeneity than strong liquids at a given block size. The maximum of $\chi_4(t)$ 
($\chi_4^P$) for fragile and solid liquids at the same relaxation time $(\tau_\alpha)$ is shown in the inset as 
a function of block size $(L_B)$. \textbf{(c)}. Dynamic  length scales as a function of rescaled temperature by 
$T_g$. Solid lines are fits to the form $\xi_d \sim |T/T_{VFT} - 1|^{-\nu}$. \textbf{(d)}. Dynamic length scales 
are plotted as a function of $|T/T_c - 1|$, where $T_c$ is the MCT divergence temperature. The 
length scale data are rescaled to make it more comparable to other data sets. The black line is a power fit with 
an exponent $\nu=0.30$ (see text for discussion). }}
  \label{fig:2}
\end{figure}

In Fig.~\ref{fig:2}(a), we show a comparison of the dynamic length scales obtained using two different 
methods for systems which reside on the two extreme ends of the spectrum in the Angell plot; first one
being the most fragile liquid with $\rho = 0.82$ (see Fig.~\ref{fig:2}(a)), while 
the other is at the strong liquid end (Fig.~\ref{fig:2}(a) inset) with $\rho = 0.65$. The quite 
good agreement between these two ways of estimations of $\xi_d$ for all state points provides us 
confidence in the measured heterogeneity length scale.
An interesting question that naturally comes up is the following: if the two extreme systems (strong and fragile) 
have very similar structural relaxation times, do they have similar dynamic heterogeneity? To investigate this we 
plot $\chi_4$ as a function of time in Fig.~\ref{fig:2}(b) for strong and fragile liquid ($\tau_\alpha$ for these 
two systems are close to each other) at roughly the same block size $(L_B)$. We find that fragile liquid show 
stronger dynamic heterogeneity than strong liquid for a given block size. In the inset of Fig.~\ref{fig:2}(b),  we 
plot the peak value of $\chi_4(t)$ as a function of $L_B$ for these two extreme systems at the same 
$\tau_\alpha$. Fragile liquid shows stronger dynamic heterogeneity at each length scale than the strong
glass-former. This observation is in stark contrast with a recent finding \cite{paul2021dynamic} of possible 
decoupling of relaxation time and dynamic heterogeneity in active glass-forming liquids. There we observed 
that heterogeneity increase with decreasing fragility if the relaxation time is kept constant. Next we focus
on the temperature dependence of $\xi_d$ with changing fragility. Fig.~\ref{fig:2}(c) shows the growth of 
dynamic length scale as a function $T_g/T$. It can be seen that for the fragile glass former, $\xi_d$ shows 
a sharp growth upon supercooling,  whereas strong glass former produces a gentle growth resembling the
Angell like plot for $\tau_\alpha$ (see Fig.\ref{fig:1}).

Inhomogeneous mode coupling theory (IMCT) \cite{PhysRevLett.97.195701}
predicts that three-point density correlation function, $\chi_3(t)$, which can be obtained by measuring
the response of the system under an external perturbation, is intimately related to the four-point 
susceptibility $\chi_4(t)$ and shows similarly scaling behaviour near the MCT transition temperature ($T_c$). 
Thus, according to IMCT, $\xi_d$ should  have a critical like behaviour: as $\xi_d \sim |T-T_c|^{-\nu}$ with
$\nu = 1/4$ being the predicted critical exponent. Recently, 
Tah et.al, \cite{PhysRevResearch.2.022067} showed for few model glass forming liquids that the 
exponent $\nu$ is in agreement with IMCT prediction for temperature near $T_c$.  
However, validity of this result across a wide range of systems with changing fragility
is not studied before.  In Fig.~\ref{fig:2}(d) we show the dependence of $\xi_d$ vs $(T-T_c)/T_c$ for 
different fragile systems. The black line  shows the power law fit to the few low temperature data points. 
We find for all the different fragile systems the value of the exponent is $\sim 0.30$ (near $T_c$) which is 
not very different from the exponent $\nu = 0.25$  expected by IMCT. 
The small difference in exponent may be 
attributable to the finite dimensions, and understanding how the exponent shifts in higher dimensions 
(e.g. near the upper critical dimension $d_c=8$) \cite{PhysRevE.81.040501,Charbonneau13939,Franz18725}, will be of considerable interest. 

\noindent{\bf Amorphous order and Fragility: }
The lack of a priori knowledge of the nature of the structural order in disorder glass forming liquids 
makes it difficult to measure the relevant degree of order. However, existing results show that 
the system has domains of different mobility near the glass transition temperature and  these 
domains have relaxation rates that are substantially faster or slower than the system's average 
relaxation rate. These different domains often called as `` cooperatively rearranging regions (CRRs)'', 
are the primary building blocks of the phenomenological Adam-Gibbs theory 
\cite{1965JChPh..43..139A} of glass transition and its subsequent development, i.e., the random 
first-order transition (RFOT) theory \cite{PhysRevA.40.1045,doi:10.1146/annurev.physchem.58.032806.104653}. 
Different mobility domains have different patch entropy, and the mean patch correlation length can 
be easily extracted from the largest groups of congruent patches which are considered to have a 
similar local order \cite{PhysRevLett.107.045501}. This hints that the origins of heterogeneous 
dynamics may be buried in the local amorphous structure. Bouchaud and Biroli 
\cite{doi:10.1063/1.1796231} suggested a non-trivial correlation function known as point-to-set (PTS) 
correlation \cite{Natphy2008} to calculate the structural or thermodynamic correlation length scale in 
supercooled glass-forming liquids in an order-agnostic way as envisaged in RFOT theory.  
Growth of this static length scale gives a notion of emerging thermodynamic order that  may 
be connected  to the dynamical slowing down of the system, even when traditional structural 
features (e.g. ``pair correlation function'') are blind to capture the dramatic slowing down of the 
relaxation time. Here, we address the crucial question of whether the growing thermodynamic 
amorphous order can universally explain the origin of a wide spectrum of fragility.

\begin{figure}[htbp]
\includegraphics[scale=0.26]{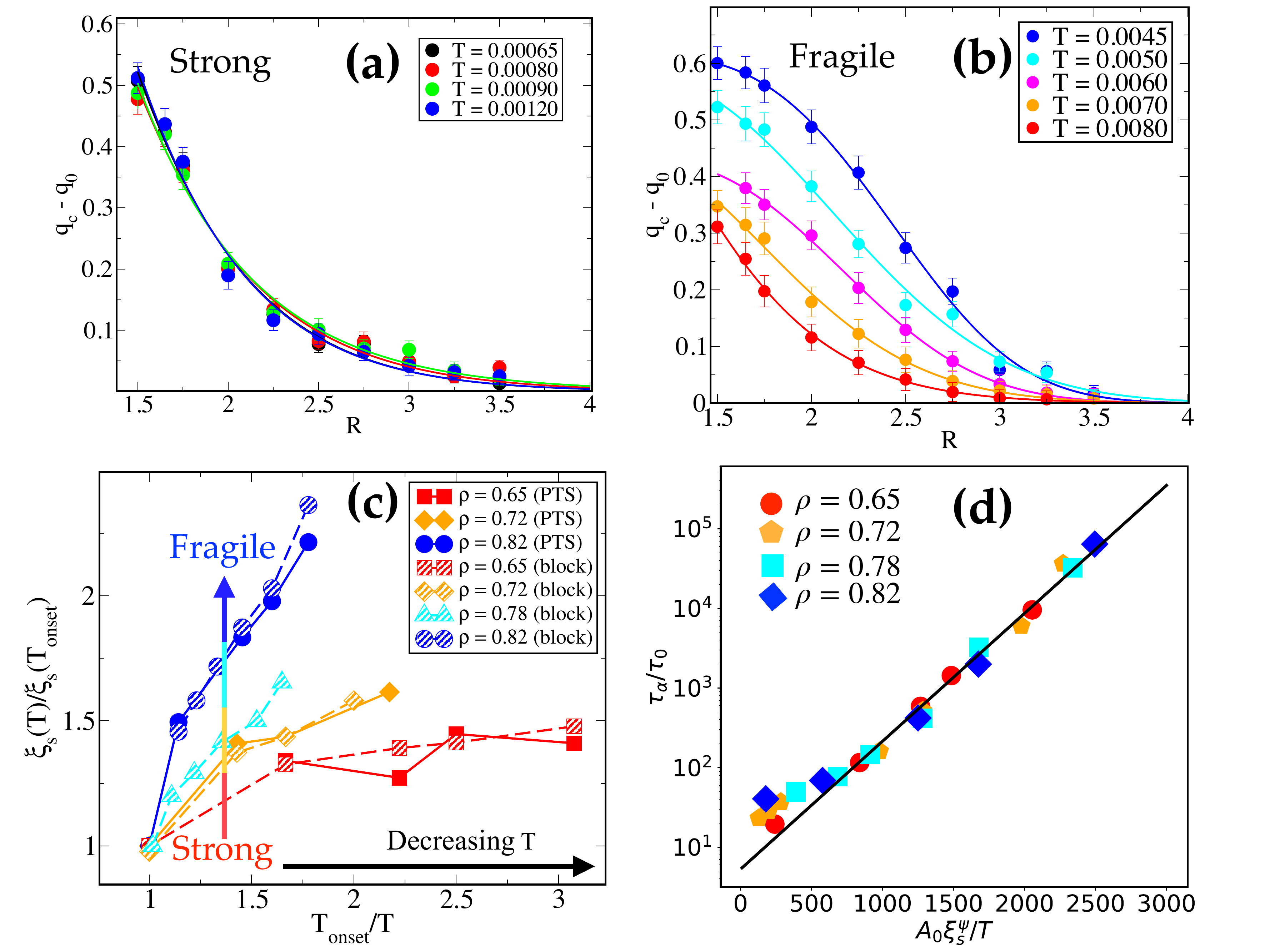}
\caption{Overlap correlation as a function of cavity radius $R$  for strong glass former \textbf{(a)} and  fragile glass former \textbf{(b)} at various temperatures. \textbf{(c)} Rescaled static length scale ($\xi_s$) as a function of reduced temperature (re-scaled by onset temperature $(T_{onset})$). Here, $\xi_s(T_{onset})$ is the value of static length scale at onset temperature. \textbf{(d)} Rescaled relaxation time  $(\tau_\alpha/\tau_0)$ as a function of $\xi_s^\psi/T$ yielded a good data collapse for all the different fragile liquids.}
  \label{fig:3}
\end{figure}
To address this, we have computed the static length scale using point-to-set correlation 
function in cavity geometry (see the SI \cite{SI}) \cite{Natphy2008,PhysRevLett.108.225506,
PhysRevLett.121.085703,Nagamanasa,PhysRevE.94.032605}. Fig.~\ref{fig:3}(a) and 
\ref{fig:3}(b) show the nature of the correlation function with increasing cavity radius, $R$
for various studied temperatures for a strong and a fragile liquid. The correlation functions for 
the strong liquid decay at the same rate for all temperatures, indicating no growth in the amorphous 
order (see Fig.~\ref{fig:3}(a)) with decreasing temperature. However, for fragile liquid these 
correlation functions show significantly slower decay as a function of $R$ with decreasing 
temperature. The estimation of the length scale $(\xi_s)$ can be done by simply fitting the 
correlation functions (see SI for details). We also extracted the static length scale using block analysis 
of relaxation time as proposed in Ref.~\cite{PhysRevLett.119.205502} (see SI \cite{SI} for details). 
The temperature dependence of the length scale for different fragile  liquids determined by 
point-to-set method and block analysis method are shown and compared in Fig.~\ref{fig:3}(c). 
We find that the length scales obtained by these two methods are very similar for all the
studied systems. For better representation, we rescaled the static length scale 
by its value at onset temperature (temperature at which the thermodynamic and dynamic 
properties begin to depart from its high temperature behaviour \cite{B008749L}) and show 
the data only for temperatures below the onset temperature (see SI).
For strong glass forming liquid $\xi_s$ (``red squares'') does not grow much with 
decreasing temperature,  but for fragile glass forming liquid it (``blue circles'') grows sharply 
as system approaches the glass transition temperature.

Within the RFOT scenario, the structural relaxation time of a system is related to $\xi_s$ as
\begin{equation}
\tau_\alpha(T) = \tau_0 \exp \left [\frac{\Delta_0 \ \xi_s^\psi(T)}{T} \right] ,
\label{RFOT_EQ}
\end{equation}
where  typical free energy barrier is $\Delta(T) =  \Delta_0\ \xi_s^\psi(T)$ with $\psi$ being an apriori unknown 
scaling exponent. $\Delta_0$ and $\tau_0$ set  the energy scale and microscopic time scale of the system. 
Thus, according to Fig.~\ref{fig:3}(c), the typical free energy barrier for a strong glass former remains nearly 
constant as $\xi_s$ (red squares) stays nearly constant, but rises (blue circles) sharply for a 
fragile liquid as the temperature is lowered towards $T_g$, resulting in super-Arrhenius behaviour. 
We believe this is the first numerical evidence that conclusively suggest that $\xi_s$ which was first 
introduced in the RFOT theory to characterize the liquid's mosaic order, controls 
the relaxation process including the drastic change in fragility with changing density in supercooled liquids. 
We also believe the above results are 
very general and constitute a first clear demonstration  that order agnostic  thermodynamic order (applicable 
to all generic glass formers) is strongly correlated to fragility, rather than some specific locally preferred 
structure (LFS) \cite{doi:10.1063/1.2773716,Hallett,PhysRevX.8.011041}.

Now, if $\tau_\alpha$ is solely depended on the thermodynamics length scales as given in Eq.\ref{RFOT_EQ}, then 
one should be collapsed all the data of relaxation time for all the systems with varying fragility on 
to a master curve if plotted against $\xi_s^\psi/T$. To validate the above argument we plotted the 
re-scaled relaxation time $(\tau_\alpha/\tau_0)$ (re-scaled by time scale $(\tau_0)$ at very high temperature where $\xi_s \to 0$) as a 
function of $\Delta_0 \xi_s^\psi (T)/T$ and find that data for all the systems with different fragility collapses on to a 
straight line as shown in Fig.~\ref{fig:3} (d). We keep the exponent $\psi = 2.6$  same while varying $\Delta_0$. 
Good data collapse implies an encouraging universality between the relaxation time and the static length scale across systems
with large change in fragility. This implies that probably system's dynamic heterogeneity is tied to the static length scale \cite{PhysRevLett.121.085703,doi:10.1146/annurev-conmatphys-031113-133848}.

\vskip +0.1in
\noindent{\bf Conclusion:}
Taken together, our findings show that by fine-tuning the system's packing fraction, one might achieve a bigger change in 
kinetic fragility in soft repulsive systems for delving deeper into the relationship between their kinetic fragility and the microscopic
structural ordering. This could be easily tested experimentally by using colloidal suspensions \cite{Johan_Nature}.  
The heterogeneity length scale for fragile liquid increases significantly faster as the system approaches the glass transition temperature, 
whereas strong liquid shows much smaller increase. All liquids show MCT critical like behavior near the MCT temperature $(T_c)$, with an 
exponent $\nu=0.30$ that is close to the exponent anticipated by IMCT theory. This suggests that all supercooled liquids with a wide range 
of kinetic fragility have a universal behavior in their dynamic heterogeneity. Finally, we show that the distinct temperature dependence of the 
order agnostic thermodynamics length scale controls the temperature dependence of the structural relaxation time for all the fragile liquid 
and suggests that widely different temperature dependency of structural relaxation in various glass-forming liquids is concealed in  
local amorphous order, whose growth determines whether a system will be strong to fragile glass formers. Since the material durability is 
an intrinsic property that is tightly linked to its fragility index, we expect that our findings will be useful to better understand 
the role of microscopic structure in the emergence of fragility in future.


IT thanks  S. Ridout and A. J. Liu  for discussions. IT wants to acknowledge use of Extreme Science and Engineering Discovery Environment (XSEDE) \cite{xsede}, which is supported by 
National Science Foundation grant number TG-BIO200091. SK would like to acknowledge funding by intramural funds at TIFR Hyderabad 
from the Department of Atomic Energy (DAE). Support from Swarna Jayanti Fellowship grants DST/SJF/PSA-01/2018-19 and SB/SFJ/2019-20/05 
are also acknowledged

\textbf{Availability of Data}

The data that support the findings of this study are
available from the corresponding author upon reasonable
request.
\nocite{*}
\bibliographystyle{ieeetr}
\bibliography{Paper_fragility}

\widetext
\clearpage
\setcounter{equation}{0}
\setcounter{figure}{0}
\renewcommand{\thefigure}{S\arabic{figure}}

\section*{\textbf{Supporting Information: Kinetic Fragility Directly Correlates with the Many-body Static Amorphous Order in
Glass-Forming Liquids}}
\section{Models and Simulation Details}

\label{modelsAndSim}
We have studied a few model glass forming liquids in three 
dimensions. The model details are given below:
\vskip +0.3cm

\noindent{\textbf{3dHP Model:}} 

We have done equilibrium molecular dynamics (MD) simulations of three-dimensional systems which interpolate between finite-temperature 
glasses and hard-sphere glasses and have been studied widely in the field of jamming physics 
\citesupp{PhysRevLett.88.075507,doi:10.1146/annurev-conmatphys-070909-104045}. The system
consists of $50:50$ binary mixture of $N$ particles 
and interacting via the 
following pair-wise potential 
\begin{equation}
V_{\alpha\beta}(r) = \epsilon_{\alpha\beta} \left ( 1-\frac{r}{\sigma_{\alpha\beta}}\right)^2
\end{equation}
for $r < \sigma_{\alpha\beta}$ and $V_{\alpha\beta}(r) = 0$ otherwise. 
Here $(\sigma_{\alpha\alpha}, \sigma_{\alpha\beta}, \sigma_{\beta\beta})$ = $(1.00,\ 1.20,\ 1.40)$, and   ($\epsilon_{\alpha\alpha}, \epsilon_{\alpha\beta}, \epsilon_{\beta\beta}) = (1.0,\ 1.0,\ 1.0) .$
The packing fraction is defined 
as $\phi =   \frac{\pi}{12} (\sigma_\alpha^3+\sigma_\beta^3) \rho$ and $\rho = N/L^3$ is 
the number density. We have used system sizes of $N = 100$ to $N = 108000$ particles in our MD simulations with periodic boundary conditions. 
We performed simulations in the density range $\rho \in [0.65, 0.82]$. 
The  system shows incipient  crystallization above packing fraction $\phi = 0.924$ $(\rho = 0.94)$ \citesupp{Berthier_2009}, which is well above the area of interest in our study. 
All reported results are in thermal equilibrium which have been carefully reviewed.

\section{Self overlap correlation function}
The self overlap correlation function or two point overlap correlation function is defined as,
\begin{equation}
Q(t) = \frac{1}{N} \sum_{i=1}^{N} w\big( |\vec{r}_i(t) - \vec{r}_i(0)| \big),
\label{EQ:1}
\end{equation}
where $w(x)$ is a window function defined as $w(x) = 1$, if $x<a$ and $0$ otherwise. The window function $w(x)$ has been introduced 
to remove the decorrelation arising from particles' vibrational motions inside the local cages that are formed by their  
neighbours. In our study, we choose $a=0.3$ which corresponds to the plateau value of the mean square displacement. However, the reported 
results are not very sensitive to the particular choice of $``a"$, as long as $``a"$ does not differ significantly from the above mentioned value. The  relaxation time $\tau_\alpha$ is defined as $\langle Q(\tau_\alpha)\rangle = 1/e$, where $\langle\cdots\rangle$ refers to ensemble average.

\section{Methods to obtain dynamic length scale}
\subsection{Block analysis method}
\begin{figure*}[!htpb]
\begin{center}
\includegraphics[scale=0.45]{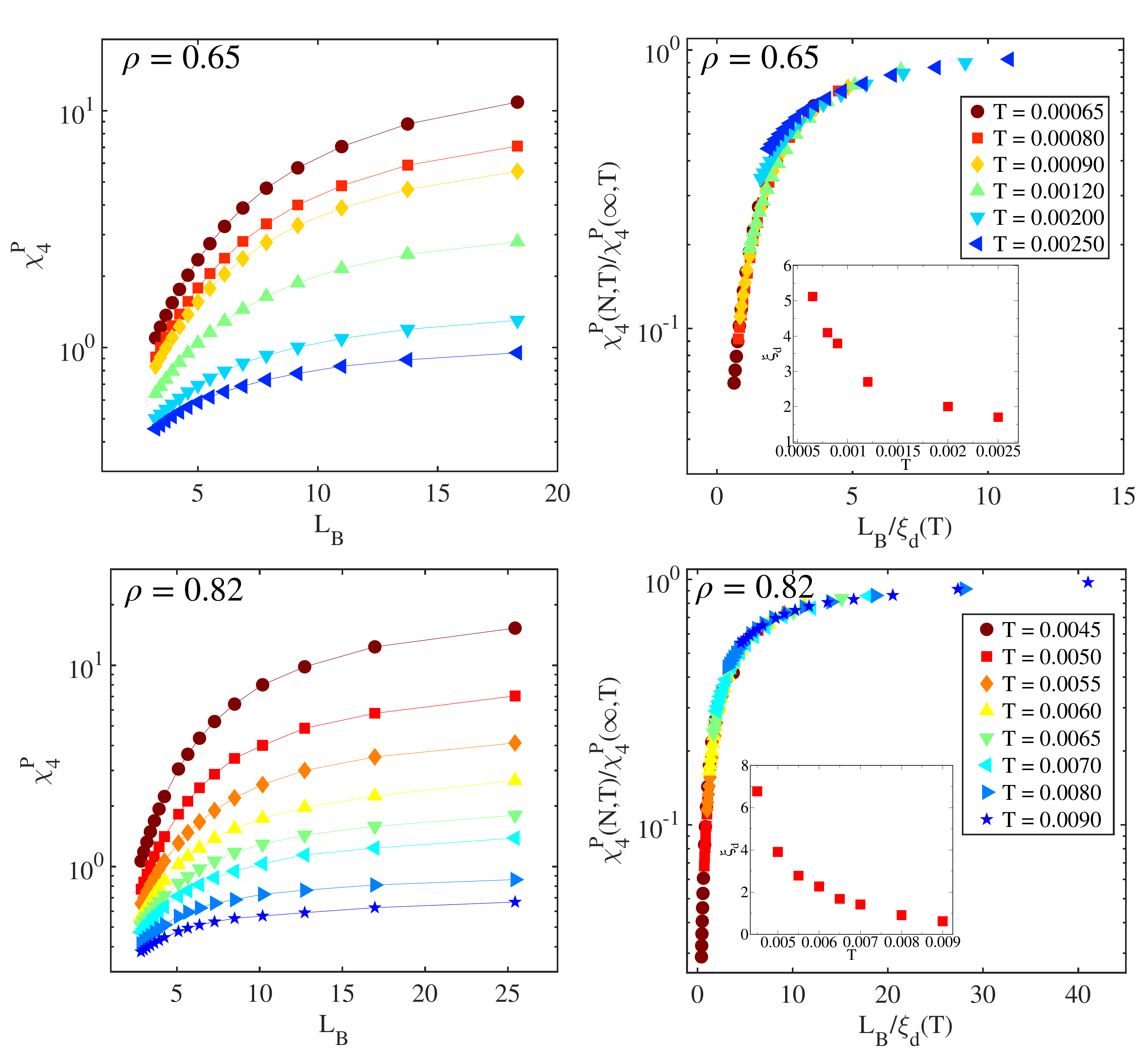}                                                                                                                
  \caption{\textbf{Top Panel:} Block size dependence of $\chi_4^P$ (left panel) and its scaling collapse (right panel) to get the dynamic length scale $(\xi_d(T))$ for strong liquid $(\rho=0.65)$. \textbf{Bottom panel:} Similar plot for fragile liquid $(\rho=0.82)$.}
  \label{Fig_SI_1}
\end{center}
\end{figure*} 
Following the procedures of Ref.~\citesupp{PhysRevLett.119.205502}, we calculated the dynamic length scale from the finite size scaling of four-point susceptibility $(\chi_4(L_B,t))$, where $L_B$ is the varying block size and $t$ is the measurement time. $\chi_4(L_B,t)$ quantifies the fluctuation of two-point correlation function $Q(L_B,t)$ 
for a given coarsening block of size $L_B$ and is defined as
\begin{equation}
\begin{split}
\chi_4(L_B,t) = N_B\left(\langle Q(L_B,t)^2\rangle - \langle Q(L_B,t)\rangle^2\right), \\
\text{and} \\
Q(L_B,t)=\frac{1}{N_B}\sum_{i=1}^{N_B}\frac{1}{n_j}\sum_{j=1}^{n_j}w(|r_j(0)-r_j(t)|).
\end{split}
\label{window}
\end{equation}
$Q(L_B,t)$ gives positional overlap between two configurations in a block of size $L_B$ separated by a time interval $t$. $N_B$ is the number of blocks of size $L_B$ and $n_j$ is the number of particles in that block at time $t=0$. In this analysis, we have evaluated the $\chi_4^P(L_B,t)$ at the time interval close to $\sim \tau_\alpha$, system's structural relaxation time, to perform the finite size scaling (FSS). Note that the system's dynamics have the maximum heterogeneity near $t \sim \tau_\alpha$. 

Now to obtain the dynamic length scale we perform finite size scaling of peak of four point dynamic susceptibility $\chi_4^P(L_B,T)$, 
which varies systematically  as the temperature drops and block size grows, implying an increasing length scale, and the data was found to be well described by the following scaling function,
\begin{equation}
\chi_4^P(L_B,T)=\chi_4^P(\infty,T)\mathcal{F}\left[\frac{L_B}{\xi_d(T)}\right],
\label{scalingAnsatz} 
\end{equation}
where $\chi_4^P(\infty,T)$  is the asymptotic value of $\chi_4^P(L_B,T)$. The dynamics length scale $(\xi_d)$ of the system can be obtained very reliably by performing data collapse to a master curve. In Fig.~\ref{Fig_SI_1} we plot the block size dependence of $\chi_4(L_B,T)$ and its scaling collapse of two extreme systems, one of which show strong liquid behaviour with $\rho = 0.65$ (top panel of  Fig.~\ref{Fig_SI_1}), while the other is at the most fragile liquid with $\rho = 0.82$ (bottom panel of Fig.~\ref{Fig_SI_1}). In the inset of Fig.~\ref{Fig_SI_1} we show the temperature dependence of the dynamics length scale for the two systems.

\subsection{Calculation of dynamical length scale from four-point structure factor $S_4(q,t)$}
\begin{figure*}[!htpb]
\begin{center}
\includegraphics[scale=0.45]{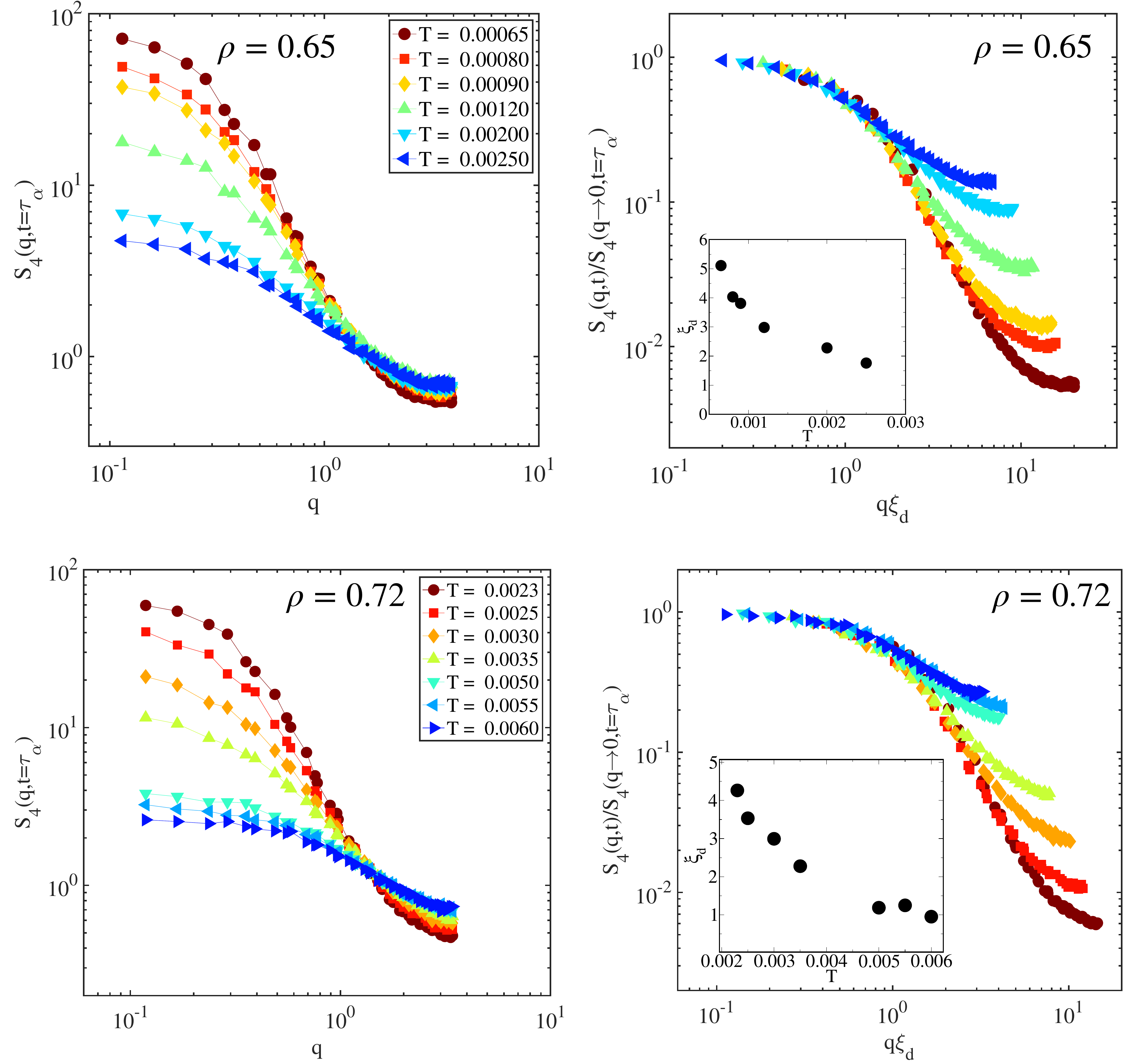}
\caption{\textbf{Top Left Panel:} Dynamic structure factor $(S_4(q,t=\tau_\alpha))$ for  $\rho=0.65$ as a function of wave vector $(q)$ for different temperature . \textbf{Top Right Panel:} Scaling plot $S_4(q,t)/S_4(q \to 0,t=\tau_\alpha)$ vs $q\xi_d$ for the 108000 particle simulations.  \textbf{Bottom Panel:} Similar plot for $\rho=0.72$. The dynamic length scale as a function of temperature  is shown in the inset.}
  \label{fig:s4qt_collapse_1}
\end{center}
\end{figure*} 

\begin{figure*}[!htpb]
\begin{center}
\includegraphics[scale=0.45]{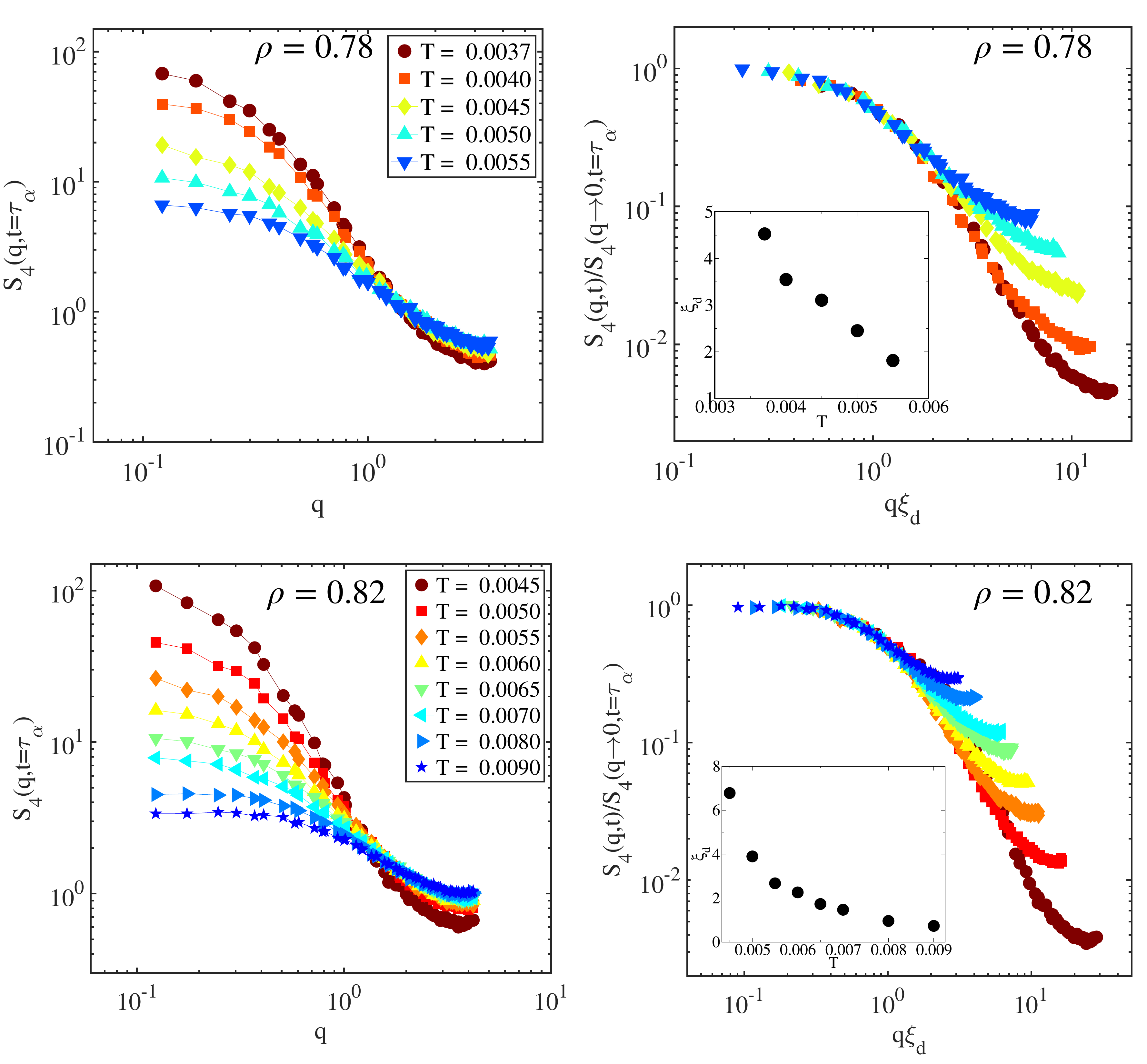}
\caption{\textbf{Top Left Panel:} Dynamic structure factor $(S_4(q,t=\tau_\alpha))$ for  $\rho=0.78$ as a function of wave vector $(q)$ for different temperature . \textbf{Top Right Panel:} Scaling plot $S_4(q,t)/S_4(q \to 0,t=\tau_\alpha)$ vs $q\xi_d$ for the 108000 particle simulations.  \textbf{Bottom Panel:} Similar plot for $\rho=0.82$. The dynamic length scale as a function of temperature  is shown in the inset.}
  \label{fig:s4qt_collapse_2}
\end{center}
\end{figure*} 
Another important quantity that can provide a lot of the information about dynamic heterogeneity 
is the four-point correlation function, $g_4(r,t)$,  and its associated susceptibilities \citesupp{chandan92} 
defined as,
\begin{equation}
\begin{split}	
g_4(r,t)= \langle \delta \rho(0,0)\delta \rho(0,t)\delta \rho(r,0)\delta \rho(r,t)\rangle -\\ \langle\delta \rho(0,0)\delta \rho(0,t)\rangle \langle\delta \rho(r,0)\delta \rho(r,t)\rangle
\end{split}	
\end{equation}
$\delta \rho(r,t)$ is the deviation of local density $\rho(r,t)$ at position $r$ and time $t$ 
from its average value $\rho_0(r,t)$, $\langle...\rangle$ represents the thermal or time 
average. The four-point time dependent structure factor $S_4(q,t)$ \citesupp{JCP119-14}, 
is related to $g_4(r,t)$ via Fourier transformation, as defined below,
\begin{equation}
S_4(q,t) = N [\langle Q(q,t)Q(-q,t)\rangle - \langle Q(q,t)\rangle ^2]
\end{equation}
where 
\begin{equation}
Q(q,t) = \frac{1}{N}\sum_{i=1}^{N}e^{i\vec{q}.\vec{r}_i(0)} 
w(|\vec{r}_i(t)-\vec{r}_i(0)|),
\end{equation}
The window function $w(x)$ is defined in the same way as it is in Eq.~\ref{window}. In the small wave vector ($q$) 
limit, the behavior of $S_4(q,t)$ can be described by the Ornstein–Zernike (OZ) form, 
\begin{equation}
S_4(q,t=\tau_\alpha) = \frac{S_4(q \to 0,t)}{1+(q\xi_d)^2}
\end{equation}
in which $\xi_d$ is the measure of the length scale that is associated with dynamically correlated region. To access the low wave vector $(q \to 0)$ limit, it is necessary to do large scale simulations. In this study, we measure the dynamics length from system that consists of $N=108000$ particles. In left panel of Fig.~\ref{fig:s4qt_collapse_1} and Fig.~\ref{fig:s4qt_collapse_2}, we plot the $S_4(q,t=\tau_\alpha)$ as a function of wave vector $q$. We fit $S_4(q,t)$ to the Ornstein-Zernike (OZ) form in the range $q\xi_d \leq 1.0$ to get the dynamic length scale.
We did a scaling plot $S_4(q,t=\tau_\alpha)/S_4(q \to 0,t=\tau_\alpha)$ vs $q\xi_d$ in the right panel of Fig.~\ref{fig:s4qt_collapse_1} and Fig.~\ref{fig:s4qt_collapse_2} using the value of $\xi_d$ obtained from the OZ fit, and we found a very good scaling collapse for all the different fragile systems. The dynamic length scale is shown as a function of temperature in the insets of all plots.

\section{Methods to obtain static length scale}
\subsection{Point-to-set method}
\label{ptsMethod}
\begin{figure*}[!htpb]
\begin{center}
\includegraphics[scale=0.32]{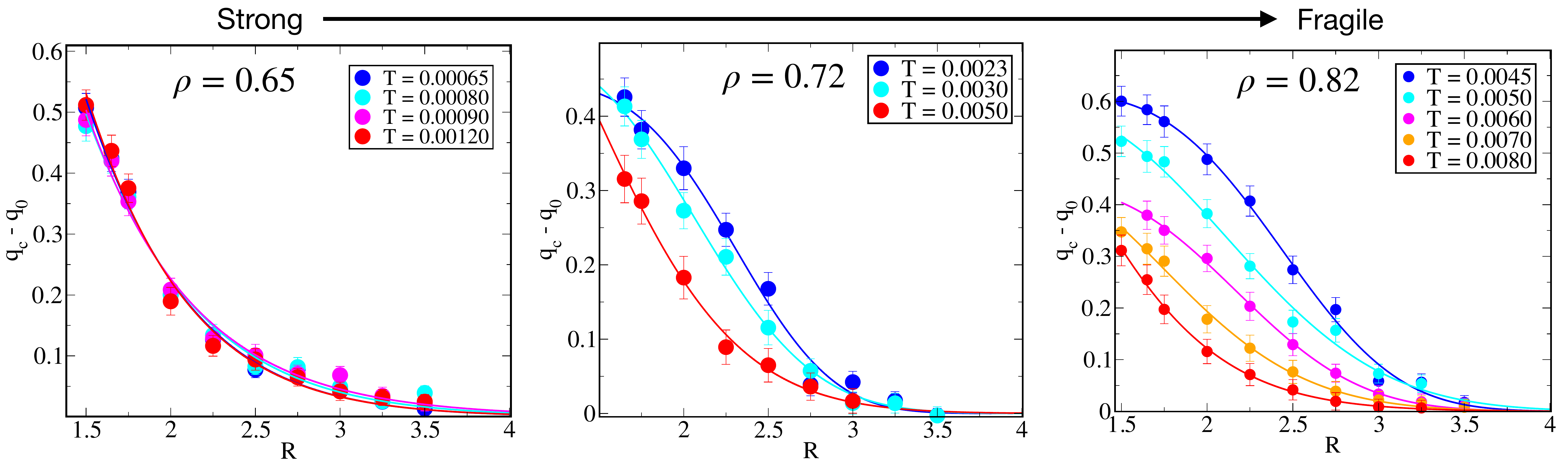}
\caption{The static overlap as a function of cavity radius $R$ for different temperature. The systems transition from strong to fragile glass (from left to right) with increasing density.}
  \label{fig:pts}
\end{center}
\end{figure*}
One of the most difficult challenges in glass physics is to determine a growing thermodynamic or 
structural  order that is correlated with the slowing down of dynamics. This becomes much 
more difficult, when conventional structural features (e.g., pair correlation function) for supercooled 
liquids and glasses don't display any noticeable differences from their high temperature liquid states. 
This problem necessitates a new approach to capturing the elusive thermodynamic order that 
emerges as the system transitions from liquid to glass while remaining disordered. To calculate 
the thermodynamic or structural length scale for amorphous systems in an order-agnostic way 
as envisioned in RFOT theory \citesupp{PhysRevA.40.1045}, Bouchaud and Biroli \citesupp{doi:10.1063/1.1796231} 
suggested a non-trivial correlation known as point-to-set (PTS) correlations \citesupp{Natphy2008}. Here, 
we follow their protocol as given in Ref.~\citesupp{Natphy2008} to measure the thermodynamic length 
scale also referred often as ``point-to-set (PTS)'' length scale. 

We perform NVT molecular dynamics simulations to produce the equilibrium bulk configurations at 
the desired density and temperature. We then start with the above equilibrium configuration and 
freeze the particles outside of a cavity of radius $R$, then re-equilibrated the particles within the 
cavity (called as mobile particles) of radius $R$ in the presence of frozen boundary conditions.  
A static overlap is established between the original equilibrated configuration and the new equilibrated 
configuration with the  frozen boundary field. To define the static overlap $(q_c(R))$, we divide the 
cavity's central region into $M$ cubic boxes with side $l$. We select the box size so that there is 
a negligible chance of having more than one particle in a box. The static overlap is defined as
\begin{eqnarray*}
q_c(R) = \lim_{t \to \infty} \frac{1}{Ml^3\rho} \sum\limits_{i=1}^{M} \langle n_i(0) n_i(t)\rangle.
\end{eqnarray*}
where $\langle..\rangle$ denotes the both thermodynamic and ensemble average. Here we have 
chosen $M = 125$ cubic boxes of side $l = 0.36$. The overlap between two similar configurations 
is $1$ after normalization and $q_0=\rho l^3$ for two totally uncorrelated configurations.  In 
Fig.~\ref{fig:pts} we plot the $q_c-q_0$ as a function of cavity radius $R$ for various temperatures 
at a given density. The length scale $(\xi_{pts})$ was calculated using a compressed exponential form 
to fit the overlap correlation as a function of radius $R$.
\begin{eqnarray*}
 \tilde{q}(R) = q_c(R) - q_0 = A\exp\left[-\left(\frac{R-a}{\xi_{pts}}\right)^\eta\right],
\end{eqnarray*}
We chose $a=1$ because cavities with $R = 1$ should only contain a single particle on average, 
and the overlap properties at this cavity size should not be affected by increasing amorphous order. 
We have shown the obtained static length scale $(\xi_s)$ in the main manuscript.

\section{Static length scale from block statistics of $\tau_\alpha$}
\begin{figure*}[!htpb]
\begin{center}
\includegraphics[scale=0.45]{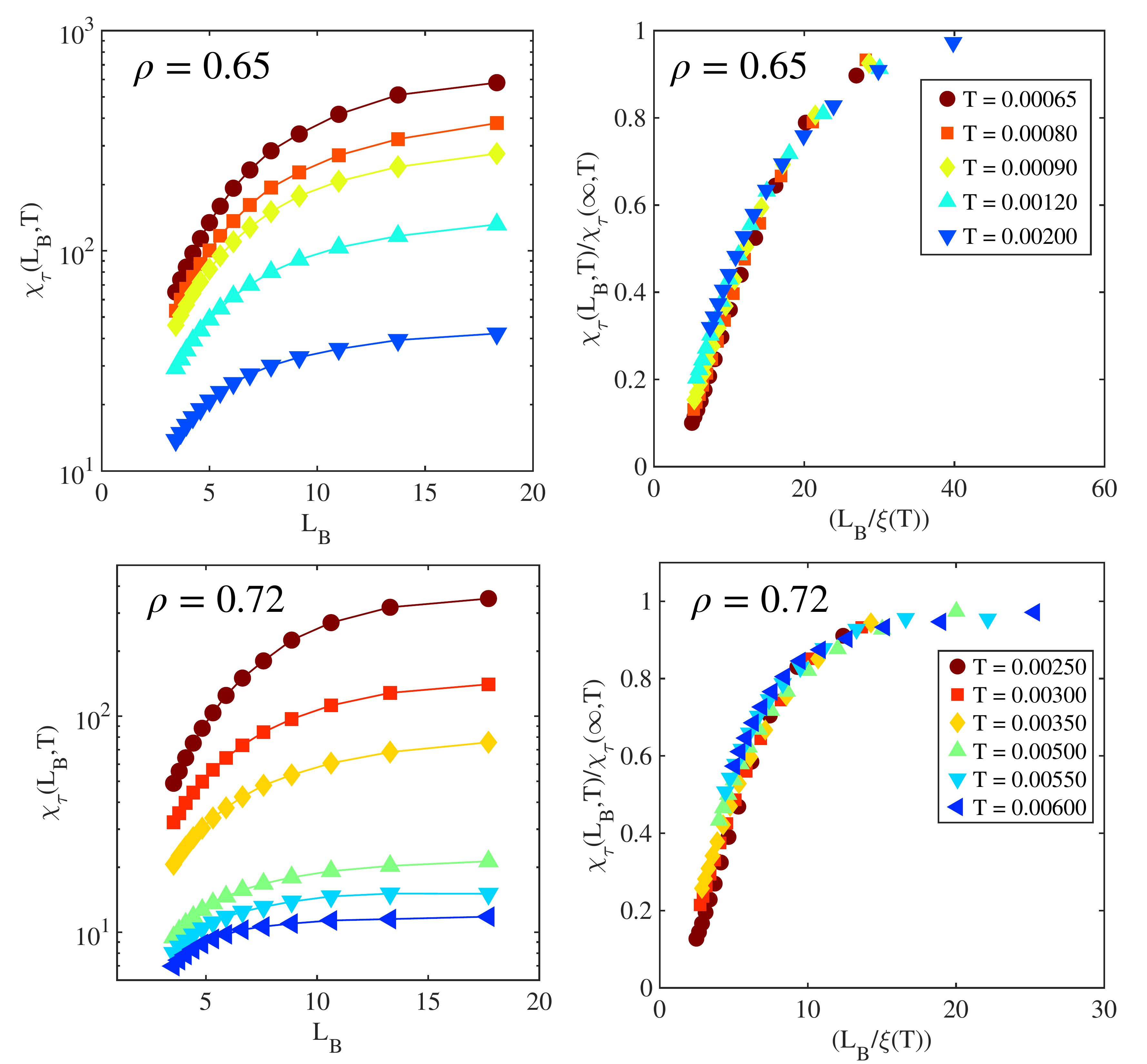}
\caption{\textbf{Top Left Panel:}  $\chi_\tau$ for  $\rho=0.65$ as a function of block size $(L_B)$  for 
different temperature. \textbf{Top Right Panel:} Scaling collapse to get the static length scale. 
\textbf{Bottom Panel:} Similar plot for $\rho=0.72$. }
  \label{fig:static_collapse_1}
\end{center}
\end{figure*} 
In Ref.~\citesupp{PhysRevLett.119.205502} an elegant and efficient method for determining the 
static length scale of the glass forming liquid from the  statistics of $\tau^{(i)}_\alpha$ on the 
block size was proposed. We have used the same procedure here. We first calculate the relaxation 
time ($\tau_\alpha^{(i)}(L_B)$) for each block using the method of Ref.~\citesupp{PhysRevLett.119.205502} 
by measuring the time at which the overlap correlation $(Q^{(i)}(L_B,t))$ for a given time origin reaches 
the value $1/e$ (the superscript $i$ indicates that the quantity is collected for a single block prior to 
any averaging).

\begin{figure*}[!htpb]
\begin{center}
\includegraphics[scale=0.45]{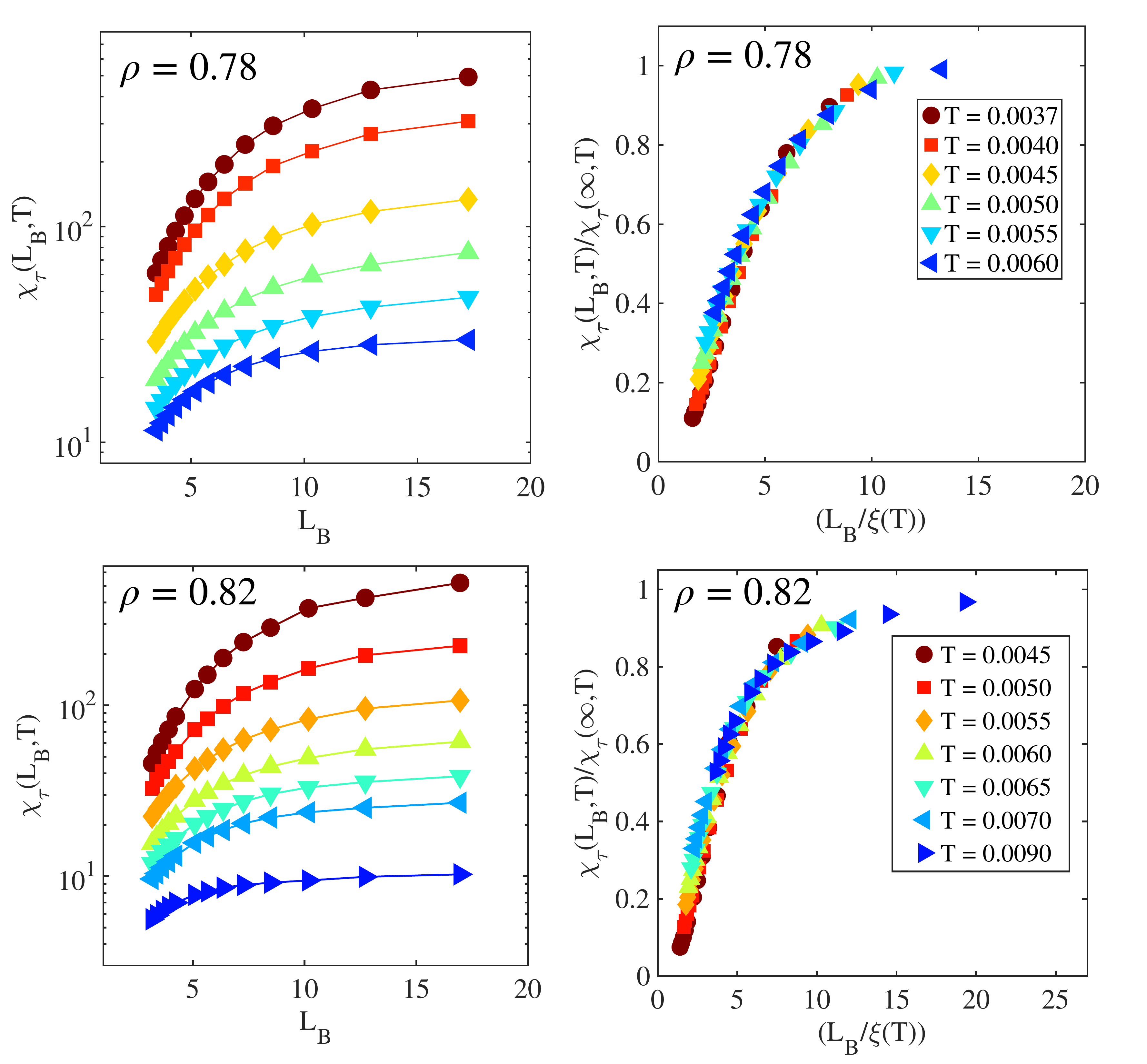}
\caption{\textbf{Top Left Panel:}  $\chi_\tau$ for  $\rho=0.78$ as a function of block size $(L_B)$  
for different temperature. \textbf{Top Right Panel:} Scaling collapse to get the static length scale. 
\textbf{Bottom Panel:} Similar plot for $\rho=0.82$. }
  \label{fig:static_collapse_2}
\end{center}
\end{figure*} 
We measured the mean and variance of the quantity  $\tau_\alpha^{(i)}(L_B)$ and define $\chi_\tau(L_B,T)$ as
\begin{equation}
\chi_\tau(L_B,T) =
L_B^3\left\langle
\frac{\frac{1}{N_B}\sum_{i=1}^{N_B}[\Delta\tau_{\alpha}^{(i)}(L_B)]^2}{{[\overline{ \tau_{\alpha}^{(i)}(L_B)}]}^2}\right\rangle,
\label{chitau}
\end{equation}
where $\overline{\tau_{\alpha}^{(i)}(L_B)} = 
\frac{1}{N_B}\sum_{i=1}^{N_B} \tau_{\alpha}^{(i)}(L_B)$, $\Delta \tau_\alpha^{(i)}(L_B) = \tau_\alpha^{(i)}(L_B)-\overline{\tau_{\alpha}^{(i)}(L_B)}$, and the time-origin averaging is denoted by the outermost angular brackets. This quantity $\chi_\tau$ has a strong block size $(L_B)$ dependence, indicating a growing length scale as temperature decreases.
We used scaling analysis to determine the static length scale that can be derived from the scaling collapse of $\chi_\tau$.
In Fig.~\ref{fig:static_collapse_1} and Fig.~\ref{fig:static_collapse_2}  we plot the $\chi_\tau$ as a function of $L_B$ and its scaling collapse to obtain the static length scale.  In the main paper, a comparison of the static length scale obtained from statistics of $\tau^{(i)}_\alpha$ on block size and point-to-set length is shown.

\section{STATIC LENGTH SCALE FROM FINITE SIZE SCALING OF $\tau_\alpha$}
\begin{figure*}[!htpb]
\begin{center}
\includegraphics[scale=0.25]{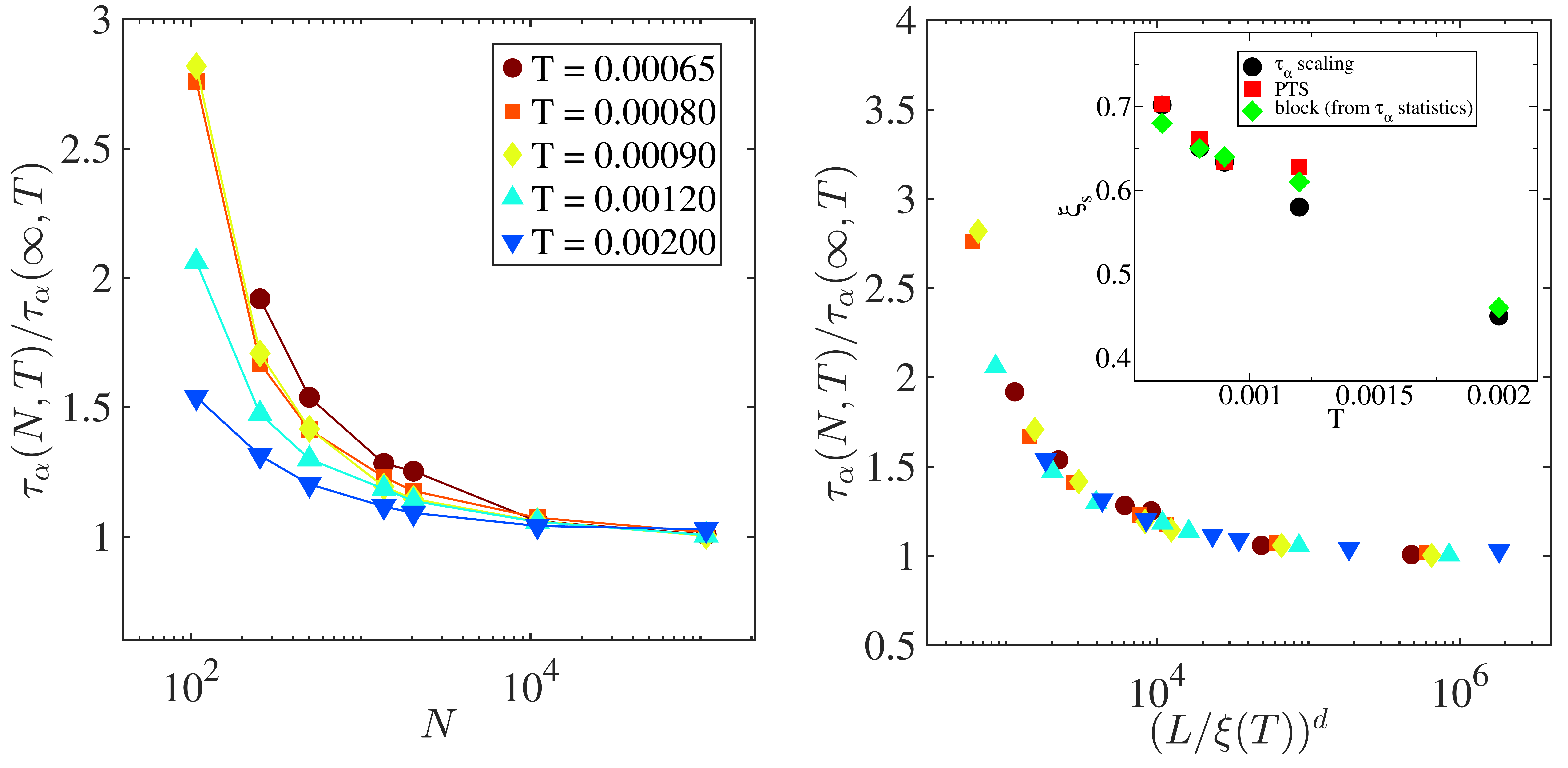}
\caption{\textbf{Left Panel:} System size dependence of $\alpha$-relaxation time  for strong liquid ($\rho=0.65$). \textbf{Right Panel:} Scaling collapse to get the static length scale. \textbf{Inset:} Comparison of the length scales from different methods. }
  \label{fig:tau_alpha_collapse}
\end{center}
\end{figure*}

Finally, we also have calculated the static length scale from finite size scaling of $\alpha$-relaxation time 
$(\tau_\alpha)$ for the strong liquid following the procedure of Ref.~\citesupp{PhysRevE.86.061502}. This is mainly
done to make sure that the growth of static length scale for strong liquids is weak as obtained using PTS and 
block analysis methods. In the left panel of Fig.~\ref{fig:tau_alpha_collapse} we  plot the re-scaled relaxation 
time (rescaled by asymptotically large system size value of the relaxation time ($\tau_\alpha(\infty,T)$) ) for 
different temperatures and system sizes. In the right panel we show the full data collapse by using the static 
length scale at different temperatures. The credibility of the obtained static length scale is bolstered by the  
reasonable data collapse.  The static length scale produced from the finite size scaling of $\tau_\alpha$, 
point-to-set method, and statistics of $\tau^{(i)}_\alpha$ on block size at different temperatures are 
compared in the inset and we find that the lengths measured using these various methods are very similar. 

\section{Calculation of Onset temperature}
\begin{figure*}
\begin{center}
\includegraphics[scale=0.47]{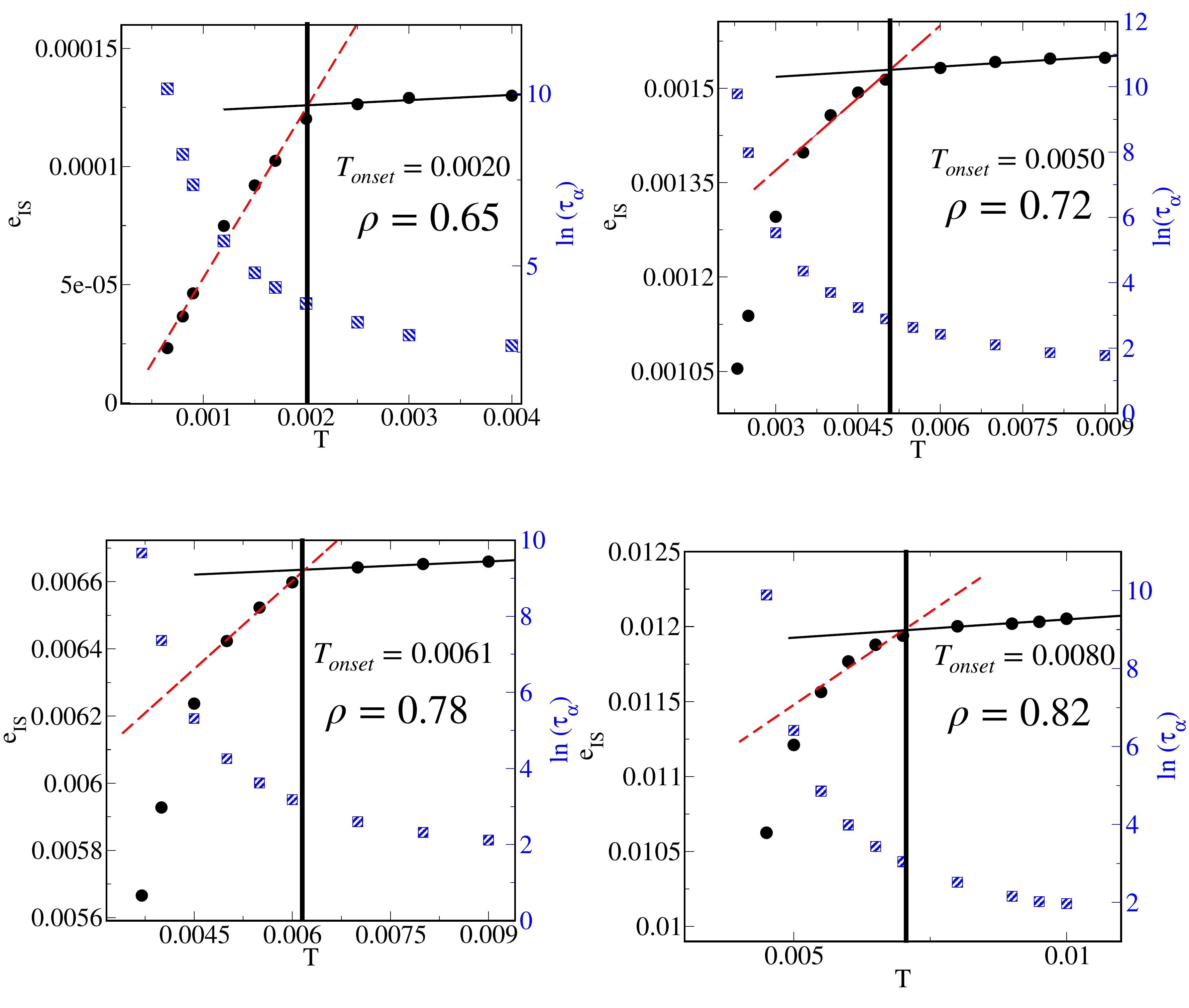}
\caption{Temperature dependence of the average inherent state energies ($e_{IS}$) shown by black circles in  left y-axis  for different fragile liquids. System's relaxation time against temperature are plotted in right y-axis. Onset temperature indicated in each plot by a vertical line.}
  \label{fig:E_IS}
\end{center}
\end{figure*}

We looked at the features of local potential energy minima sampled by the liquid to find a crossover 
or onset temperature below which the liquid starts to show very sluggish dynamics and thermodynamic 
and dynamical traits begin to depart from its high temperature liquid \citesupp{sastry_nature,B008749L,sri_Nature}. 
This approach in the field of disordered system is well-known as the energy landscape view or paradigm. 
To get the inherent state (IS), we performed energy minimization over $1000$ configurations at each  
temperature and density  state point. Inherent states for different fragile liquids are obtained for a wide 
range of temperatures, from high to very close to the glass transition point. In left y-axis of Fig.~\ref{fig:E_IS}, 
we plot the average IS energies ($e_{IS}$, black circles) against each of the studied temperatures for 
different fragile liquids. The plot shows that for high temperature, the average inherent state energy is 
nearly temperature independent and begins to show a strong temperature dependence at a certain 
crossover temperature.  The deviation point from the high temperature is taken as the crossover 
temperature or onset temperature of ``slow dynamics'' (indicated in each plot by a vertical line). On the 
right y-axis of Fig.~\ref{fig:E_IS} we plot the relaxation time (blue squares) as a function of temperature. 
We show that the system's relaxation time begins to increase from the same onset temperature as mined 
by the inherent state energy analogy. Above the onset temperature, the system relaxes in an exponential 
manner, similar to a high-temperature liquid. We have used these onset temperatures to scale the data 
of the temperature dependence of static length scale for a good comparison.
\clearpage
\bibliographystylesupp{ieeetr}
\bibliographysupp{Paper_fragility} 

\end{document}